\documentclass[12pt]{article}
\usepackage{graphicx}
\begin{document}
\begin{titlepage}

\title{The teleparallel equivalent of general relativity and the gravitational
centre of mass}

\author{J. W. Maluf$\,^{(\ast)}$\\
Instituto de F\'{\i}sica, \\
Universidade de Bras\'{\i}lia\\
C. P. 04385 \\
70.919-970 Bras\'{\i}lia DF, Brazil\\}
\date{}
\maketitle
\begin{abstract}
We present a brief review of the teleparallel equivalent of general
relativity and analyse the expression for the centre of mass density of the 
gravitational field. This expression has not been sufficiently discussed in
the literature. One motivation for the present analysis is the investigation of
the localization of dark
energy in the three-dimensional space, induced by a cosmological constant in a 
simple Schwarzschild-de Sitter space-time. We also investigate the gravitational
centre of mass density in a particular model of dark matter, in the space-time 
of a point massive particle and in an arbitrary space-time with axial symmetry.
The results are plausible, and lead to the notion of gravitational centre of 
mass (COM) distribution function.
\end{abstract}
\thispagestyle{empty}
\vfill
\noindent PACS numbers: 04.20.Cv, 04.20.-q, 04.70.Bw\par
\bigskip
\noindent $(\ast )$ wadih@unb.br, jwmaluf@gmail.com\par
\end{titlepage}
\newpage

\section{Introduction}
The most popular and acceptable approach to the relativistic theory of 
gravitation is given by Einstein's theory of general relativity. However,
nowadays there are several alternative formulations of theories for the
gravitational field that attempt to explain the dark energy and dark matter
problems, which do not find satisfactory explanations within the framework of 
Einstein's general relativity. Moreover, concepts such as energy, momentum,
angular momentum and centre of mass of the gravitational field are usually 
defined only for asymptotically flat space-times, in the context of a 3+1 type
formulation. The latter are definitions for the total quantities,
and suffer of at least two restrictions: the definitions are valid only for
asymptotically flat space-times, and there do not exist localized expressions
for the densities of the energy-momentum and 4-angular momentum of the 
gravitational field. The ADM definition for the gravitational energy-momentum
\cite{ADM} is constructed out of the metric tensor, and by means of the metric
tensor it is not possible to construct suitable scalar densities in the form of
total divergences. The approach via pseudo-tensors is certainly not 
satisfactory. The notions of energy-momentum and angular momentum of the 
gravitational field have been extensively discussed in the literature, but not
the concept of gravitational centre of mass. 

The notion of centre of mass can be made clear in flat space-time. 
Any relativistic field theory in flat space-time is expected to be covariant 
under the inhomogeneous Lorentz transformations, or Poincar\'e 
transformations: the 4-rotations and space-time translations. The generators of
these transformations satisfy an algebra, the algebra of the Poincar\'e group.
The generators
of the 4-rotations are composed by the generators of the ordinary 3 dimensional
rotations, and by the generators of the boosts. The latter are related to the 
centre of mass moment of the field. Energy, momentum and angular momentum of the
field constitute 7 conserved integral quantities associated to the 
symmetries of the theory. The integrals are carried out over the whole 
three-dimensional space. The 3 other integral quantities are associated to
the centre of mass of the field, which sometimes is also called the centre of 
energy \cite{Weinberg}. 

In the notation of Ref. \cite{Weinberg}, the centre of mass integrals read

\begin{equation}
J^{0i}=tP^i-\int d^3x\,x^iT^{00}\,,
\label{1}
\end{equation}
where $0$ and $i$ are time and space indices, 
$P^i$ is the i-th component of the momentum of the field, and $T^{00}$ is
the energy component of the energy-momentum tensor of field. It is argued 
\cite{Weinberg} that the $J^{0i}$ components have no clear physical significance
since $J^{0i}$ can be made to vanish if the coordinate system is chosen to 
coincide with the ``centre of energy'' at $t=0$. However, in the context of
general relativity, Dixon \cite{Dixon1,Dixon2,Dixon3} developed a procedure for 
describing the dynamics of extended bodies in an arbitrary gravitational field,
and for this  purpose a definition of the centre of mass of
such bodies (considered as quasi-rigid bodies) was proposed. A 
general relation between the centre of mass 4-velocity and the energy-momentum
of the body was obtained \cite{Ehlers}. One is led to the concept of centre 
of mass world line, whose uniqueness depends on the strength of the 
gravitational field \cite{Ehlers}.

In the standard metric formulation of general relativity, the centre of mass 
moment for the gravitational field has been first considered by Regge and
Teitelboim \cite{Regge1,Regge2}, and reconsidered by several other authors (see 
Refs. \cite{Beig,Baskaran,Nester1,Nester2} and references therein). The centre of 
mass integral was obtained in the context of the Hamiltonian formulation of 
general relativity. The idea was to require the variation of the total Hamiltonian
to be well defined in an asymptotically flat space-time, where the standard 
asymptotic space-time translations and 4-rotations are considered as coordinate
transformations at spacelike infinity. 
This requirement leads to the addition of boundary (surface) terms
to the primary Hamiltonian, so that the latter has well defined functional
derivatives, and therefore one may obtain the field equations in the Hamiltonian
framework (Hamilton's equations) by means of a consistent procedure.  In this 
way, one arrives at the total energy, momentum, angular momentum and centre of 
mass moment of the gravitational field, given by surface terms of the total 
Hamiltonian.

In this article we address the centre of mass moment of the gravitational field
in the realm of the teleparallel equivalent of general relativity (TEGR), which 
is an alternative and mathematically consistent formulation of general relativity 
\cite{Maluf1} (see also Ref. \cite{Pereira}, and chapters 5 and 6 of  Ref.
\cite{Blagojevic} and references therein). The geometrical structure of the TEGR
was already considered by Einstein \cite{Einstein1,Einstein2} in his attempt to 
unify gravity and electromagnetism, and later on by Cho \cite{Cho1,Cho2}, Hayashi 
and Shirafuji \cite{Hayashi1,Hayashi2}, 
Hehl et. al. \cite{Hehl10}, Nitsch \cite{Nitsch}, 
Schweizer et. al. \cite{Schweizer}, Nester \cite{Nester} and Wiesendanger 
\cite{Wiesendanger}). In recent years the teleparallel geometrical structure has
been used in modified theories of gravity, with the purpose of constructing 
cosmological models that provide a consistent explanation to the dark energy
problem (see the review article \cite{Cai} and references therein).

The TEGR is constructed 
out of the tetrad fields $e^a\,_\mu$, where $a=\lbrace (0),(i)\rbrace$ and 
$\mu=\lbrace 0,i\rbrace$ are SO(3,1) and space-time indices, respectively.
The extra 6 components of the tetrads (compared to the 10 components of the 
metric tensor) yield additional geometric structure, that allows to define
field quantities that cannot be constructed in the ordinary metric formulation
of the theory (such as non-trivial total divergences, for instance). The tetrad 
fields allow to use concepts and definitions of both Riemannian and 
Weitzenb\"ock geometries.

The definitions of the gravitational energy, momentum, angular momentum and 
centre of mass moment in the TEGR are not obtained according to the procedure 
described above, based on surface integrals of the total Hamiltonian. 
In the TEGR we first consider the Hamiltonian 
formulation of the theory \cite{Maluf2,Maluf3}. The constraint equations of the
theory (typically as $C =0$) are equations that define the energy-momentum and
the 4-angular momentum of the gravitational field \cite{Maluf1} (i.e.,
$C=H-E =0$). Moreover, the 
definitions of the energy-momentum and 4-angular momentum satisfy the algebra
of the Poincar\'e group in the phase space of theory \cite{Maluf1,Maluf4}.
However, the energy-momentum definition, together with the gravitational 
energy-momentum tensor (but not the 4-angular momentum) may also be
obtained directly from the Lagrangian field equations \cite{Maluf1}.

The gravitational centre of mass moment to be considered here yields the
concept of gravitational centre of mass (COM) distribution function.
One purpose of the present article is to show that a cosmological constant,
which might be responsible for the dark energy, induces a very intense 
(divergent) gravitational COM distribution function in the vicinity of
the cosmological horizon $r=R \simeq\sqrt{3/\Lambda}$ in a simple 
Schwarzschild-de Sitter space-time, in agreement with the hypothetical existence
of dark energy. It seems that this result, obtained by means of tetrad fields,
cannot be obtained in the context of the metric formulation of general 
relativity.

In Section 2 we present a brief review of the TEGR, emphasizing a recent 
simplified definition of the 4-angular momentum of the gravitational field,
given by a total divergence. In Section 3 we investigate the gravitational COM
distribution function of (i) the space-time of
a massive particle in isotropic coordinates, (ii) the Schwarzschild-de Sitter
space-time, (iii) a particular model of dark energy that arises from the
non-local formulation of general relativity, and (iv) of an arbitrary space-time
with axial symmetry. In the analysis of the first three cases above, which are
spherically symmetric, we arrive at interesting results, that share similarities
with the standard expressions in classical mechanics. For such space-times,
the total centre of mass moment vanishes, as expected.
\bigskip

\noindent {\bf Notation}:
space-time indices $\mu, \nu, ...$ and SO(3,1) (Lorentz) indices
$a, b, ...$ run from 0 to 3. The torsion tensor is given by 
$T_{a\mu\nu}=\partial_\mu e_{a\nu}-\partial_\nu e_{a\mu}$. The flat space-time 
metric tensor raises and lowers tetrad indices, and is fixed by 
$\eta_{ab}= e_{a\mu} e_{b\nu}g^{\mu\nu}=(-1,+1,+1,+1)$.
The frame components are given by the inverse tetrads 
$\lbrace e_a\,^\mu \rbrace$. The determinant of the tetrad fields is written as
$e=\det(e^a\,_\mu)$.

It is important to note that we assume that the space-time geometry is 
determined by the tetrad fields only, and thus the only
possible non-trivial definition for the torsion tensor is given by
$T^a\,_{\mu\nu}$. This tensor is related to the 
antisymmetric part of the Weitzenb\"ock  connection 
$\Gamma^\lambda_{\mu\nu}=e^{a\lambda}\partial_\mu e_{a\nu}$, which
determines the Weitzenb\"ock space-time and the distant parallelism of vector
fields.

\section{A review of the Lagrangian and Hamiltonian formulations of the
TEGR}

The TEGR is constructed out of the tetrad fields only. The first relevant 
consideration is an identity between the scalar curvature and an invariant 
combination of quadratic terms in the torsion tensor, 

\begin{equation}
eR(e) \equiv -e\left({1\over 4}T^{abc}T_{abc} + 
{1\over 2}T^{abc}T_{bac} - T^{a}T_{a}\right)
+ 2\partial_{\mu}(eT^{\mu})\,,
\label{2}
\end{equation}
where $ T_{a} = T^{b}\,_{ba}$ and 
$T_{abc} = e_{b}\,^{\mu}e_{c}\,^{\nu}T_{a\mu\nu}$.
The Lagrangian density for the gravitational field in the TEGR is given by 
\cite{Maluf5}

\begin{eqnarray}
L(e) &=& -k\,e\left({1\over 4}T^{abc}T_{abc} + {1\over 2}T^{abc}T_{bac} 
- T^{a}T_{a}\right) - {1\over c}L_{M}\nonumber \\
& \equiv & -ke\Sigma^{abc}T_{abc} - {1\over c}L_{M}\,,
\label{3}
\end{eqnarray}
where $k = c^3/(16\pi G)$, $L_{M}$ represents the Lagrangian density for the 
matter fields, and $\Sigma^{abc}$ is defined by
\begin{equation}
\Sigma^{abc} = {1\over 4}\left(T^{abc} + T^{bac} - T^{cab}\right) 
+ {1\over 2}\left(\eta^{ac}T^{b} - \eta^{ab}T^{c}\right)\,.
\label{4}
\end{equation}
Thus, the Lagrangian density is geometrically equivalent to the scalar curvature
density. The variation of $ L(e)$ with respect to $e^{a\mu}$
yields the fields equations 

\begin{equation}
e_{a\lambda}e_{b\mu}\partial_\nu (e\Sigma^{b\lambda \nu} )-
e (\Sigma^{b\nu}\,_aT_{b\nu\mu}-
{1\over 4}e_{a\mu}T_{bcd}\Sigma^{bcd} )={1\over {4kc}}eT_{a\mu}\,,
\label{5-1}
\end{equation}
where $T_{a\mu}$ is defined by 
${{\delta L_M}/ {\delta e^{a\mu}}}=eT_{a\mu} $.
The field equations are equivalent to Einstein's equations. It is possible to
verify by explicit calculations that the equations above can be rewritten as 

\begin{equation}
\label {5-2}
{1\over 2}\lbrack R_{a\mu}(e) - {1\over 2}e_{a\mu}R(e)\rbrack
={1\over {4kc}}\,T_{a\mu}\,,
\end{equation}

Since the Lagrangian density (\ref{3}) does not contain the total divergence 
that arises on the right hand side of Eq. (\ref{2}), it is not invariant under
arbitrary local SO(3,1) transformations, but the field equations (\ref{5-1}) are
covariant under such transformations. 

The equivalence between the TEGR and the standard metric formulation of general
relativity is based on the equivalence of Eqs. ({\ref{5-1}) and ({\ref{5-2}). 
However, in the TEGR there are additional field quantities (like third order
tensors) constructed by means of the tetrad fields, such as total divergences, 
for instance, that cannot be obtained in the standard metric formulation. 
These additional field quantities are covariant under the global Lorentz 
transformations, but not under local  
transformations. In the ordinary formulation of arbitrary field theories, 
energy, momentum, angular momentum and COM moment are frame dependent field 
quantities, that transform under the global SO(3,1) transformations. In 
particular, energy transforms as the zero component of the energy-momentum
four-vector. This feature must hold also in the presence of the 
gravitational field. As an
example, consider the total energy of a black hole, represented by the mass
parameter $m$. As seen by a distant observer, the total energy of a static
Schwarzschild black hole is given by $E=mc^2$. However, at great distances the 
black hole may be considered as a particle of mass $m$, and if it moves with 
constant velocity $v$, then its total energy as seen by the same distant 
observer is $E=\gamma mc^2$, where $\gamma = (1-v^2/c^2)^{-1/2}$.  Likewise, 
the gravitational momentum, angular momentum and the COM moment are also frame
dependent field quantities in general, whose values are different for different
frames and different observers. On physical grounds, energy, momentum, angular 
momentum and COM moment cannot be local Lorentz {\it invariant} field 
quantities, since these quantities depend on the frame, as we know from special
relativity, which is the limit of the general theory of relativity when the 
gravitational field is weak or negligible. 

After some rearrangements, Eq. (\ref{5-1}) may be written in the form 
\cite{Maluf1}

\begin{equation}
\partial_{\nu}(e\Sigma^{a\mu\nu}) = 
{1\over 4k}ee^{a}\,_{\nu}(t^{\mu\nu} + {1\over c}T^{\mu\nu})\,,
\label{5}
\end{equation}
where 
\begin{equation}
t^{\mu\nu} = k(4\Sigma^{bc\mu}T_{bc}\,^{\nu} - g^{\mu\nu}\Sigma^{bcd}T_{bcd})\,,
\label{6}
\end{equation}
is interpreted as the gravitational energy-momentum tensor \cite{Maluf1,Maluf6}
and $T^{\mu\nu} = e_{a}\,^{\mu}T^{a\nu}$. 

The Hamiltonian density of the TEGR is constructed as usual in the phase space 
of the theory. We first note that the Lagrangian density (\ref{3}) does not 
depend on the time derivatives of $e_{a0}$. Therefore, the latter arise as 
Lagrange multipliers in the Hamiltonian density $H$. The momenta canonically 
conjugated to $e_{a0}$ are denoted by $\Pi^{a0}$. The latter are primary 
constraints of the
theory: $\Pi^{a0} \approx 0$. The momenta canonically conjugated to $e_{ai}$ are
given by $\Pi^{ai} = \delta L/\delta \dot{e}_{ai}= -4k\Sigma^{a0i}$.
The Hamiltonian density is obtained by rewriting the 
Lagrangian density in the form $L = \Pi^{ai}\dot{e}_{ai} - H$,
in terms of $e_{ai}, \Pi^{ai}$ and Lagrange multipliers. After the Legendre
transform is performed, we obtain the final form of the Hamiltonian density.
It reads \cite{Maluf4,Maluf3}

\begin{equation}
H(e,\Pi) = e_{a0}C^{a} + \lambda_{ab}\Gamma^{ab}.
\label{7}
\end{equation}
where $\lambda_{ab}$ are Lagrange multipliers. In the above equation we have 
omitted a surface term. $C^{a} = \delta H/\delta e_{a0}$  is a long 
expression of the field variables, and  $\Gamma^{ab}=-\Gamma^{ba}$ are defined 
by

\begin{equation}
\Gamma^{ab} = 2\Pi^{[ab]} + 4ke(\Sigma^{a0b} - \Sigma^{b0a})\,.
\label{8}
\end{equation}
After solving the field equations, the Lagrange multipliers are identified as
$\lambda_{ab} = (1/4)(T_{a0b}-T_{b0a}+e_{a}\,^{0}T_{00b}-e_{b}\,^{0}T_{00a})$.
The constraints $C^{a}$ may be written as

\begin{equation}
C^{a} = -\partial_{i}\Pi^{ai} - p^{a} = 0\,,
\label{9}
\end{equation}
where $p^{a}$ is an intricate expression of the field quantities. 

The quantities
$C^{a}$ and $\Gamma^{ab}$ are first class constraints. They satisfy an algebra 
similar to the algebra of the Poincar\'e group \cite{Maluf3}. The integral form
of the constraint equations $C^{a} = 0$ yield the gravitational energy-momentum
$P^{a}$ \cite{Maluf1},
\begin{equation}
P^{a} = - \int_{V}d^{3}x\,\partial_{i}\Pi^{ai}\,,
\label{10}
\end{equation}
where $V$ is an arbitrary volume of the three-dimensional space and 
$\Pi^{ai} = -4k\Sigma^{a0i}$. In similarity to the definition above, the 
definition of the gravitational 4-angular momentum follows from the constraint
equations $\Gamma^{ab}=0$ \cite{Maluf4}. However, it has been noted \cite{Maluf7}
that the second term on the right hand side of Eq. (\ref{8}) can be rewritten as
a total divergence, so that the constraints $\Gamma^{ab}$ become

\begin{equation}
\Gamma^{ab} = 2\Pi^{[ab]} - 2k\partial_{i}[e(e^{ai}e^{b0} -
 e^{bi}e^{a0})] = 0\,.
\label{11}
\end{equation}
Therefore, the definition of the total 4-angular momentum of the gravitational
field $L^{ab}$ may be given by an integral of a total divergence, in 
similarity to Eq. (\ref{10}). We have 

\begin{equation}
L^{ab} = -\int_{V}d^{3}x\,2\Pi^{[ab]}\,,
\label{12}
\end{equation}
where 
\begin{equation}
2\Pi^{[ab]} =(\Pi^{ab} - \Pi^{ba}) 
= 2k\partial_{i}[e(e^{ai}e^{b0} - e^{bi}e^{a0})]\,.
\label{13}
\end{equation}

It is easy to show \cite{Maluf4} that expressions (\ref{10}) and (\ref{12}) 
satisfy the algebra of the Poincar\'e group in the phase space of the theory,

\begin{eqnarray}
\{P^{a}, P^{b}\}& =&0\,, \nonumber \\
\{P^{a},L^{bc}\}&=& \eta^{ab}P^{c} - \eta^{ac}P^{b}\,,  \nonumber \\
\{L^{ab},L^{cd}\} &=& \eta^{ad}L^{cb} + \eta^{bd}L^{ac} - 
\eta^{ac}L^{db} - \eta^{bc}L^{ad}\,.
\label{14}
\end{eqnarray}
Therefore, from a physical point of view, the interpretation of the 
quantities $P^{a}$ and $L^{ab}$ is consistent.

Definitions (\ref{10}) and (\ref{12}) are invariant under coordinate 
transformations of the three-dimensional, under time reparametrizations, and
under global SO(3,1) transformations. The gravitational energy is the zero 
component of the energy-momentum four vector $P^a$.

\section{The centre of mass moment}

The gravitational centre of mass (COM) moment is given by the components

\begin{equation}
L^{(0)(i)}=-\int d^3x\,M^{(0)(i)}\,,
\label{15}
\end{equation}
where

\begin{equation}
M^{(0)(i)}=2\Pi^{[(0)(i)]}=2k\partial_j\lbrack e(e^{(0)j}e^{(i)0}-
e^{(i)j}e^{(0)0})\rbrack\,,
\label{16}
\end{equation}
according to definition (\ref{13}). The quantity $-M^{(0)(i)}$ is
identified as the gravitational COM density. The evaluation of the expression
above is very simple. One needs just to establish the suitable set of tetrad 
fields that define a frame in space-time.

The inverse tetrads $e_a\,^\mu$ are 
interpreted as a frame adapted to a particular class of observers in space-time.
Let the curve $x^\mu(\tau)$ represent the timelike worldline $C$ 
of an observer in space-time, where $\tau$ is the proper time of the observer.
The velocity of the observer along $C$ is given by $u^\mu=dx^\mu/d\tau$.
A frame adapted to this observer is constructed by identifying the
timelike component of the frame $e_{(0)}\,^\mu$ with the velocity $u^\mu$
of the observer: $e_{(0)}\,^\mu=u^\mu(\tau)$. The three other 
components of the frame, $e_{(i)}\,^\mu$, are orthogonal to $e_{(0)}\,^\mu$, and
may be oriented in the
three-dimensional space according to the symmetry of the physical system.
If the space-time has axial symmetry, for instance, then the $e_{(3)}\,^\mu$
components of the tetrad fields are chosen to be oriented, asymptotically,  
along the $z$ axis of the coordinate system, i.e., 
$e_{(3)}\,^\mu(t,x,y,z)\simeq(0,0,0,1)$ in the limit $r\rightarrow \infty$.
A static observer in space-time is defined by the condition $u^\mu=(u^0,0,0,0)$.
Thus, a frame adapted to a static observer in space-time must satisfy the
conditions $e_{(0)}\,^i(t,x^k)=(0,0,0)$. 

An alternative way to characterise a frame in space-time is by means of the 
acceleration tensor $\phi_{ab}$ \cite{Mashh2,Mashh3,Maluf81,Maluf82,Maluf83},

\begin{equation}
\phi_{ab}={1\over 2}\lbrack T_{(0)ab}+T_{a(0)b}-T_{b(0)a}\rbrack\,.
\label{17}
\end{equation}
This tensor is invariant under coordinate transformations and covariant under
global SO(3,1) transformations, but not under
local SO(3,1) transformations. It yields the inertial (i.e.,  the 
non-gravitational) accelerations that are necessary to impart to a frame in
space-time in order to maintain the frame in a given inertial state.  Three
components of $\phi_{ab}$ yield the translational accelerations, and three 
other components yield the frequency of rotation of the frame. 
Altogether, these six
components cancel the gravitational acceleration, so that the frame is 
kept in a particular inertial state.

In the following, we will evaluate the density of the centre of mass moment of
four space-time configurations that exhibit spherical symmetry. In the four 
cases we will establish the frame of a static observer in space-time.

\subsection{The space-time of a massive point particle}

The Schwarzschild solution in isotropic coordinates represents the space-time of
a point massive particle \cite{Parker, Katanaev}. It is obtained as an exact 
solution of Einstein's equations by writing the energy-momentum tensor in terms
of a $\delta$ function of a point particle of mass $M$, with support at the 
origin of the coordinate system. The solution is described by the line element

\begin{equation}
ds^2=-\alpha^2 c^2 dt^2+\beta^2\lbrack dr^2+r^2(d\theta^2+
\sin^2\theta\,d\phi^2)\rbrack\,,
\label{18}
\end{equation}
where 

\begin{equation}
\alpha^2=\biggl( { { 1-{m \over {2r}}\over {1+{m\over {2r}}}}}\biggr)^2\,,
\ \ \ \ \ \ 
\beta^2= \biggl( 1+ {m\over {2r}}\biggr)^4\,.
\label{19}
\end{equation}
The parameter $m=GM/c^2$ represents the mass of the point particle that appears 
in the energy-momentum tensor. The line element above is clearly a solution of
Eq. (\ref{5-2}), with the appropriate energy-momentum tensor $T_{a\mu}$ 
described in Ref. \cite{Katanaev}.

By performing a coordinate transformation to $(x,y,z)$ coordinates where

\begin{eqnarray}
x&=& r \sin\theta \cos \phi \,, \nonumber \\
y&=& r \sin\theta \sin \phi\,, \nonumber \\
z&=& r\cos\theta\,,
\label{20}
\end{eqnarray}
the  line element becomes

\begin{equation}
ds^2= -\alpha^2\,c^2 dt^2+\beta^2(dx^2+dy^2+dz^2)\,.
\label{21}
\end{equation}
The tetrad fields adapted to static observers is given by

\begin{eqnarray}
e_{a\mu}(t,x,y,z)=\pmatrix{-\alpha&0&0&0\cr
0&\beta&0&0\cr
0&0&\beta &0\cr
0&0&0&\beta\cr}\,.
\label{22}
\end{eqnarray}
Taking into account Eq. (\ref{16}), 
straightforward calculations yield $M^{(0)(1)}=2k\,\partial_1 \beta^2$, 
$M^{(0)(2)}=2k\,\partial_2 \beta^2$ and $M^{(0)(3)}=2k\,\partial_3 \beta^2$.
It is easy to obtain

\begin{eqnarray}
-M^{(0)(1)}&=&d_{g}\,x \,, \nonumber \\
-M^{(0)(2)}&=&d_{g}\,y \,, \nonumber \\
-M^{(0)(3)}&=&d_{g}\,z \,.
\label{23}
\end{eqnarray}
The quantity $d_{g}$ is defined by 
\begin{equation}
d_{g}={{4\,k\,m}\over r^3}\biggl(1+{m\over {2r}}\biggr)^3\,.
\label{24}
\end{equation}
Therefore,

\begin{eqnarray}
L^{(0)(1)}&=&\int d^3x\, d_{g}x \,, \nonumber \\
L^{(0)(2)}&=&\int d^3x\, d_{g}y \,, \nonumber \\
L^{(0)(3)}&=&\int d^3x\, d_{g}z \,,
\label{25}
\end{eqnarray}
where $d^3x=dx\,dy\,dz$ and $r^2=x^2+y^2+z^2$. The expressions above remind
the definition of centre of mass in classical mechanics. Given that 
$M^{(0)(i)}=2k\,\partial_i \beta^2$, it is easy to see that all integrals given
by Eq. (\ref{15}) vanish, namely, all components of the total centre of mass 
moment vanish. However, the field quantity (\ref{24}) has the following 
properties: 

\begin{eqnarray}
r \rightarrow \infty\,&:& \ \ \ \ d_{g} \rightarrow 0\,, \nonumber \\
r \rightarrow 0 \,&:& \ \ \ \ d_{g} \rightarrow \infty\,.
\label{26}
\end{eqnarray}
Thus, $d_{g}$ is more intense in the vicinity of the particle, and vanishes 
at spatial infinity. In view of Eqs. (\ref{25}) and (\ref{26}), $d_{g}$ may be
interpreted as the gravitational COM distribution function. It is clearly 
related to the intensity of the gravitational field.  The analyses of the 
space-time configurations below support this interpretation, as we will see.

\subsection{The Schwarzschild-de Sitter space-time}

The line element of the Schwarzschild-de Sitter space-time is given by

\begin{equation}
ds^2=-\alpha^2\,dt^2+{1\over \alpha^2}dr^2 + r^2d\theta^2+
r^2\sin^2\theta d\phi^2\,,
\label{27}
\end{equation}
where

\begin{equation}
\alpha^2=1-{{2m}\over r} -{{r^2}\over {R^2}}\,,
\label{28}
\end{equation}
$R=\sqrt{3/\Lambda}$ and $\Lambda$ is the cosmological constant. Here we are
considering the speed of light $c=1$. The Schwarzschild-de Sitter space-time has
been considered in the TEGR in Ref. \cite{Maluf9}. The set of tetrad fields 
adapted to stationary observers in space-time is given by

\begin{equation}
e_{a\mu}=\pmatrix{-\alpha&0&0&0\cr
0&\alpha^{-1}\sin\theta\,\cos\phi&r\cos\theta\,\cos\phi
&-r\sin\theta\,\sin\phi\cr
0&\alpha^{-1}\sin\theta\,\sin\phi&r\cos\theta\,\sin\phi
&r\sin\theta\,\cos\phi\cr
0&\alpha^{-1}\cos\theta&-r\sin\theta&0\cr}\,.
\label{29}
\end{equation}
After long but simple calculations we find that the components of Eq. (\ref{16})
read

\begin{eqnarray}
-M^{(0)(1)}&=& 4k\sin\theta \biggl({1\over \alpha}-1\biggr)\,
r\sin\theta \cos\phi\,, \nonumber \\
-M^{(0)(2)}&=& 4k\sin\theta \biggl({1\over \alpha}-1\biggr)\,
r\sin\theta \sin\phi\,, \nonumber \\
-M^{(0)(3)}&=& 4k\sin\theta \biggl({1\over \alpha}-1\biggr)\,
r\cos\theta\,.
\label{30}
\end{eqnarray}
We identify $x=r\sin\theta\cos\phi$, $y=r\sin\theta\sin\phi$, $z=r\cos\theta$
as usual, and write Eqs. (\ref{15}) as

\begin{eqnarray}
L^{(0)(1)}&=& \int d^3 x\,4k\sin\theta\biggl({1\over \alpha}-1\biggr)
\,x\,,\nonumber \\
L^{(0)(2)}&=& \int d^3 x\,4k\sin\theta\biggl({1\over \alpha}-1\biggr)
\,y\,,\nonumber \\
L^{(0)(3)}&=& \int d^3 x\,4k\sin\theta\biggl({1\over \alpha}-1\biggr)
\,z\,,
\label{31}
\end{eqnarray}
where $d^3x=dr\,d\theta\,d\phi$. Integration in the angular variables implies 
the vanishing of the three integrals $L^{(0)(i)}$, i.e., the total centre of
mass vanishes, as expected. The equations above may be
written exactly as Eq. (\ref{25}) provided we identify

\begin{equation}
d_g=4k\,\sin\theta \biggl({1\over \alpha}-1\biggr)\equiv
4k\,\sin\theta\, f(r)\,.
\label{32}
\end{equation}

The analysis of the expression above leads to interesting results. Let $r_1$
and $r_2$ denote the two horizons of the Schwarzschild-de Sitter space-time, 
$\alpha(r_1)=0$ and $\alpha(r_2)=0$, so
that $r_1<r_2$. The radius $r_1$ is close to the Schwarzschild radius, 
$r_1\approx {{2m}\over r}$, and $r_2\approx R$.  We have

\begin{eqnarray}
r \rightarrow r_1 \,&:& \ \ \ \ f(r) \rightarrow \infty\,, \nonumber \\
r \rightarrow r_2 \,&:& \ \ \ \ f(r) \rightarrow \infty\,.
\label{33}
\end{eqnarray}
The function $f(r)$ is defined by Eq. (\ref{32}).
The minimum of $f(r)$ is given by 

$${{df}\over {dr}}=-{1\over \alpha^2} {{d\alpha}\over {dr}}=0\,,$$
and takes place at
$r_{min}=(mR^2)^{1/3}$. Thus, $d_g$ is intense close to both $r_1$ and $r_2$,
i.e., close to the Schwarzschild and cosmological horizons. 

The radial position $r_{min}$ is related to the inertial accelerations of an 
observer. In order to understand this feature, we evaluate the translational 
(non-gravitational) accelerations of a frame given by Eq. (\ref{17}). We find

\begin{eqnarray}
\phi_{(0)(1)}&=&{{d\alpha}\over{dr}}\sin\theta\cos\phi\,,\nonumber \\
\phi_{(0)(2)}&=&{{d\alpha}\over{dr}}\sin\theta\sin\phi\,,\nonumber \\
\phi_{(0)(3)}&=&{{d\alpha}\over{dr}}\cos\theta\,.
\label{34}
\end{eqnarray}
We define the inertial acceleration vector $\Phi$ as 

\begin{equation}
\Phi(r)=(\phi_{(0)(1)}, \phi_{(0)(2)},\phi_{(0)(3)})\equiv\phi(r) \hat{r}
={{d\alpha}\over{dr}} \hat{r} \,,
\label{35}
\end{equation}
where $\hat{r}=(\sin\theta\cos\phi, \sin\theta\sin\phi, \cos\theta)$. Since

$${{d\alpha}\over {dr}}={1\over \alpha}\biggl({m\over r^2} - 
{r\over R^2}\biggr)\,,$$
we see that 

\begin{eqnarray}
r_1<r<r_{min}: \ \ \ \ {{d\alpha}\over {dr}}>0 \ \ \ 
\rightarrow \ \ \ \phi(r)>0\,,
\nonumber \\
r_{min}<r<r_2: \ \ \ \ {{d\alpha}\over {dr}}<0 \ \ \ 
\rightarrow \ \ \  \phi(r)<0\,.
\label{36}
\end{eqnarray}
Thus, given that the inertial acceleration $\phi(r)>0$ is repulsive in the
region $r_1<r<r_{min}$, the gravitational acceleration is attractive in this
interval, as expected. By means of a similar argument, we see that the 
gravitational acceleration is repulsive in the region $r_{min}<r<r_2$, as
expected.

In view of the analysis above, we may interpret $d_g$ given by Eq. (\ref{32}) 
as the gravitational COM distribution function, in similarity to Eq. (\ref{24}),
and therefore one may understand the gravitational repulsion as attraction to a
region of intense gravitational COM distribution function, which, in the
present case, is the region in the vicinity of the cosmological horizon. If
dark energy is indeed related to the existence of a cosmological constant, 
then it is natural that it is concentrated close to the radius 
$r_2\approx R=\sqrt{3/\Lambda}$ in the context of a simple 
Schwarzschild-de Sitter model.

The function $d_g$ plays the role of a gravitational COM density. However,
mathematically it is not a density. The integrands in Eqs. (\ref{25}) and
(\ref{31}) are in fact densities, but not $d_g$ alone. In Newtonian 
mechanics, $d_g$ in Eqs. (\ref{25}) and (\ref{31}) plays the role of mass
density.

\subsection{Dark matter simulated by non-local gravity}

A non-local formulation of general relativity, based on a geometrical framework
similar to the one established by Eqs. (\ref{2},\ref{3},\ref{4}) has been 
developed by Hehl, Mashhoon and collaborators \cite{B1,B2,B3}. One interesting
consequence of this development is an extension of Newtonian gravity that may 
play a relevant role in the dynamic of galaxies, and might provide an 
explanation that is expected to come from dark matter models of gravity. 
We restrict the considerations to a simplified space-time with spherical
symmetry, so that Eqs. (\ref{27}), (\ref{29}), (\ref{31}) and (\ref{32}) remain
valid. 

The Newtonian approximation is established by 

\begin{equation}
\alpha^2 = -g_{00} \simeq 1+ {{2\Phi_g}\over c^2}\,,
\label{37}
\end{equation}
where $\Phi_g$ is the Newtonian potential, and $2\Phi_g/c^2<<1$.
It follows that 

\begin{equation}
f(r)={1\over \alpha} -1 \simeq  -{\Phi_g \over c^2}\,.
\label{38}
\end{equation}
The Newtonian potential that arises in the non-local formulation of gravity is
given by \cite{B2,B3}

\begin{equation}
\Phi_g \simeq-{{GM}\over r}+{{GM}\over \lambda}
\ln\biggl({r\over \lambda}\biggr)\,,
\label{39}
\end{equation}
where $\lambda$ is a constant length, and is taken to be 
$\lambda\approx 1kpc=3260$ light-years. Consequently, the influence of the 
second term on the right hand side of Eq. (\ref{39}) in the solar system is 
negligible. Therefore, we find  

\begin{equation}
f(r)={1\over \alpha}-1\simeq {m\over r}-{m\over \lambda}
\ln \biggl( {r\over \lambda}\biggr)\,.
\label{40}
\end{equation}
In the expression above, $m=GM/c^2$.  For values of $r$ within a galaxy, 
$r<\lambda$ and thus 
$-(m/ \lambda) \ln ( r/ \lambda)$ is positive, and decreases as $1 /r $
with increasing values of $r$, a result that shows that the gravitational field
is sufficiently intense at the borders of a galaxy to explain the rotation 
curves of spiral galaxies. The function $d_g=4k\,\sin\theta f(r)$ may again
be understood as the gravitational COM distribution function of the spherically 
symmetric space-time.

\subsection{Arbitrary space-time with axial symmetry}

The analysis of a space-time that is not spherically symmetric allows to obtain
the generalization of Eqs. (\ref{25}) and (\ref{31}). Let us consider an 
arbitrary space-time with axial symmetry. It  is described by following line 
element,

\begin{equation}
ds^2=g_{00}dt^2+g_{11}dr^2+g_{22}d\theta^2+g_{33}d\phi^2
+2g_{03}d\phi\, dt\,,
\label{41}
\end{equation}
where all metric components depend on $r$ and $\theta$, but not on $\phi$ :
$g_{\mu\nu}=g_{\mu\nu}(r,\theta)$. The determinant $e=\sqrt{-g}$ is 
$e= \lbrack g_{11}g_{22}\,\delta \rbrack^{1/2}$, where 

$$\delta= g_{03}g_{03}-g_{00}g_{33}\,.$$
The inverse metric components are 
$g^{00}=- g_{33} /\delta$, 
$g^{03}=g_{03}/ \delta$ and
$g^{33}= - g_{00} /\delta$.
The set of tetrad fields in spherical coordinates that is adapted to static 
observers in space-time is given by

\begin{equation}
e_{a\mu}=\pmatrix{-A&0&0&-C\cr
0&\sqrt{g_{11}} \,\sin\theta \cos\phi&
\sqrt{g_{22}} \cos\theta \cos\phi & -D\, r \sin\theta \sin\phi\cr
0& \sqrt{g_{11}}\, \sin\theta \sin\phi&
\sqrt{g_{22}} \cos\theta \sin\phi &  D\, r \sin\theta \cos\phi\cr
0& \sqrt{g_{11}}\, \cos\theta & -\sqrt{g_{22}}\sin\theta&0}\,.
\label{42}
\end{equation}
The functions $A, C$ and $D$ are defined such that Eq. (\ref{42}) yields 
(\ref{41}). They read

\begin{eqnarray}
A(r,\theta)&=& (-g_{00})^{1/2}\,, \nonumber \\
C(r,\theta)&=&-{{g_{03}}\over{(-g_{00})^{1/2}}}\,, \nonumber \\
D(r,\theta)
&=&{1\over (r\sin\theta)} \biggl[ {{\delta} \over {(- g_{00})}}
\biggr]^{1/2} \,.
\label{43}
\end{eqnarray}
After simple calculations, we find that Eqs. (\ref{15}) and (\ref{16}) yield

\begin{eqnarray}
L^{(0)(1)}&=&\int d^3x\, d_{g1}\,(r\sin\theta\cos\phi) \,, \nonumber \\
L^{(0)(2)}&=&\int d^3x\, d_{g2}\,(r\sin\theta\cos\phi) \,, \nonumber \\
L^{(0)(3)}&=&\int d^3x\, d_{g3}\,(r\cos\theta) \,,
\label{44}
\end{eqnarray}
where now we have

\begin{eqnarray}
d_{g1}=d_{g2}&=&2k\biggl\{-{1\over r}\partial_1\biggl[
{{g_{22}\delta}\over {(-g_{00})}} \biggr]^{1/2} - {1\over {r\sin\theta}}\,
\partial_ 2 \biggl[\biggl( {{g_{11}\delta }\over {(-g_{00})}}\biggr)^{1/2}
\cos\theta  \biggr] \nonumber \\
&{}&+ {1\over {r\sin\theta}}(g_{11}g_{22})^{1/2} \biggr\}\,, \nonumber \\
d_{g3}&=&2k\biggl\{ -{1\over r}\partial_1 \biggl[
{{g_{22}\delta}\over {(-g_{00})}} \biggr]^{1/2}+
{1\over {r\cos\theta}}\,
\partial_ 2 \biggl[\biggl( {{g_{11}\delta }\over {(-g_{00})}}\biggr)^{1/2}
\sin\theta \biggr] \biggr\} \,.
\label{45}
\end{eqnarray}
In the flat space-time, the quantities above vanish. It is not difficult to see
that if the metric tensor components above represent the exterior gravitational
field of a typical rotating source, the expressions above are not divergent.
Note that $L^{(0)(1)}$ and $L^{(0)(2)}$ vanish due to integration in $\phi$, as
a consequence of the axial symmetry, but $L^{(0)(3)}$ is non-vanishing in 
general.

In the equations above we obtain $d_{g1}=d_{g2} $ because of the axial
symmetry of the space-time. We see that, in general, we may have three different
COM distribution functions, one for each direction in the 
three-dimensional space, in contrast to the situation in classical mechanics, 
where there is a single mass density in the definition of centre of mass.

\section{Concluding remarks}

In this article we have investigated the definition of centre of mass of the
gravitational field, in the realm of the teleparallel equivalent of general
relativity. The analysis of the gravitational centre of mass density leads to
the concept of COM distribution function. We may understand the latter as a
quantity that provides a description of the intensity of the gravitational
field in space-time. The emergence of this quantity justifies the analysis of
the centre of mass density of arbitrary configurations of the gravitational 
field, including gravitational wave configurations. We have applied this 
definition to the space-time endowed with a positive cosmological constant. 
At the speculative level, dark energy might be a consequence of the existence 
of a positive cosmological constant that induces a strong gravitational 
acceleration very far from our present location in the universe. In the simple
model established by the Schwarzschild-de Sitter space-time, dark energy is 
roughly located in the region beyond $r=r_{min}=(mR^2)^{1/3}$, according to 
Eq. (\ref{36}).

The centre of mass moment naturally arises in the 
Hamiltonian formulation of the teleparallel equivalent of general relativity,
and its definition is obtained from the primary constraints of the theory - 
Eq. (\ref{11}). It is given by Eqs. (\ref{15}) and (\ref{16}). The analysis led
us to interpret the quantity $d_g$ in the integrand of Eqs. (\ref{25}), 
(\ref{31}) and (\ref{44}) as the gravitational COM distribution function. 
Although $d_g$ plays the role of a density, mathematically it is not a density.
It vanishes when the gravitational field is turned off. The 
expressions of $L^{(0)(i)}$ given by Eqs. (\ref{25}) and (\ref{31}) do
remind the standard expression of centre of mass in classical mechanics. The 
distribution function $d_g$ in the three-dimensional space is related to the 
intensity of the gravitational field. In the space-time of a point massive
particle, $d_g$ is intense (and in fact diverges) in the vicinity of the 
particle, and in the Schwarzschild-de Sitter space-time $d_g$ is positive 
definite and diverges at both the Schwarzschild and cosmological horizons,
which are precisely the regions where the gravitational field is more intense.

In relativistic field theory or in the Newtonian approximation of general 
relativity, energy, momentum and angular momentum are frame dependent field 
quantities, and so they are, in general, in the present context. In particular,
the gravitational COM moment is evaluated in the frame adapted to an arbitrary 
observer in space-time. The gravitational centre of mass given by eqs. (17) and
(18) is invariant under coordinate transformations of the three-dimensional 
space, and under time reparametrizations. It transforms covariantly under 
global SO(3,1) transformations, provided the tetrad fields transform as
$\tilde{e}^a\,_\mu=\Lambda^a\,_b e^b\,_\mu$, where $\Lambda^a\,_b$ are
matrices of the SO(3,1) group. However, definition (17) is not covariant under
local SO(3,1) transformations. In relativistic field theory, the COM definition
is also not covariant under local SO(3,1) transformations. 

We conclude that repulsion, in the Schwarzschild-de Sitter space-time, is in 
fact attraction to a region of intense gravitational COM distribution function.
We have seen that in the region $r<r_{min}= (mR^2)^{1/3}$ the gravitational 
acceleration is attractive, and is repulsive in the dark energy region 
$r>r_{min}$. We expect the present analysis to be useful in the investigation 
of realistic  cosmological models endowed with a positive cosmological 
constant.\par

\bigskip
\noindent {\bf Acknowledgement}
I am grateful to B. Mashhoon for enlightening comments and for pointing out 
relevant references.

\end{document}